\documentclass[journal,comsoc]{IEEEtran}
\usepackage{amsmath,amsfonts,amssymb}
\usepackage{algorithmic}
\usepackage{algorithm}
\usepackage[caption=false,font=normalsize,labelfont=sf,textfont=sf]{subfig}
\usepackage{stfloats}
\usepackage{url}
\usepackage{verbatim}
\usepackage{graphicx}
\usepackage{cite}
\usepackage{textcomp}
\usepackage[english]{babel}
\usepackage{multirow}    
\usepackage{booktabs}    
\usepackage{siunitx}     
\usepackage{makecell}    
\usepackage{pifont}
\usepackage{bm}
\usepackage[nolist,nohyperlinks]{acronym}
\usepackage[table]{xcolor}
\usepackage{colortbl}
\definecolor{Gray}{gray}{0.9}
\usepackage[caption=false]{subfig}
\usepackage[colorlinks=true, allcolors=black]{hyperref}  
\usepackage{array}
\usepackage{tabularx}

\usepackage{eso-pic}

\DeclareMathOperator*{\argmin}{arg\,min}
\newcommand{\HH}{\textrm{H}} 
 
\newcommand{\conj}[1]{#1^*}

\begin{acronym}
        \acro{ACF}{autocorrelation function}
        \acro{AWGN}{additive white Gaussian noise}
        \acro{CFR}{channel frequency response}
        \acro{DFT}{discrete time Fourier transform}
        \acro{EPDF}{empirical \ac{PDF}}
        \acro{FM}{frequency modulation}
	\acro{GEV}{generalized extreme value}        
        \acro{LS}{least squares}
        \acro{LTI}{linear time invariant}	
	\acro{LPTV}{linear periodically time-varying}		
        \acro{MIMO}{multiple-input multiple-output}
        \acro{ML}{maximum likelihood}
        \acro{MPM}{multipath propagation model}
        \acro{NRMSE}{normalized root mean square error}	
        \acro{OFDM}{orthogonal frequency division multiplexing}
        \acro{PDF}{probability density function}        
        \acro{PHY}{physical layer}
	\acro{PL}{path loss}
        \acro{PLC}{power line communication}
        \acro{PSD}{power spectral density}
	\acro{RMSE}{root mean square error}
        \acro{RV}{random variable}
	\acro{SISO}{single-input-single-output}
	\acro{WLS}{weighted \ac{LS}}	
\end{acronym}

\definecolor{reviewer1}{rgb}{0 0 0}
\definecolor{reviewer2}{rgb}{0 0 0}
\definecolor{reviewer3}{rgb}{0 0 0}

\begin{document}
\AddToShipoutPicture{%
  \put(0,\LenToUnit{\paperheight-1.8cm}){%
    \makebox[\paperwidth][c]{%
      \footnotesize
      \parbox{\textwidth}{
\vspace{-2.2cm}
© 2025 IEEE.  Personal use of this material is permitted.  Permission from IEEE must be obtained for all other uses,
 in any current or future media, including reprinting/republishing this material for advertising or promotional purposes,
 creating new collective works, for resale or redistribution to servers or lists, or reuse of any copyrighted component of this work in other works.
}

    }%
  }%
}

	
	\title{A Close Examination of the Multipath Propagation Stochastic Model for Communications over Power Lines}
	
	
	\author{Jos\'e~A.~Cort\'es, Alberto~Pittolo, Irene~Povedano, Francisco~J.~Ca\~nete, and Andrea~M.~Tonello
	\thanks{Jos\'e~A.~Cort\'es, Irene~Povedano and Francisco~J.~Ca\~nete are with the Communications and Signal Processing Lab, Telecommunication Research Institute (TELMA), Universidad de Málaga, E.T.S.I. Telecomunicacion, 29010 Málaga, Spain.}
	\thanks{Alberto~Pittolo was with the University of Udine, Udine 33100, Italy.}
	\thanks{Andrea M. Tonello is with the Institute of Networked and Embedded Systems, Alpen-Adria-Universit\"{a}t Klagenfurt, Klagenfurt 9020, Austria.}
	\thanks{The work of Jos\'e~A.~Cort\'es, Irene~Povedano and Francisco~J.~Ca\~nete has been partially supported by the Spanish Government under project PID2019-109842RBI00/AEI/10.13039/501100011033 and the University of Málaga under project PAR 15/2023. The work of Alberto Pittolo and Andrea M. Tonello has been partially supported by the University of Udine and the University of Klagenfurt.}}
	
    
	\maketitle
%

	
	
\begin{abstract}
This paper focuses on the parameterization of the multipath propagation model (MPM) for indoor broadband power line communications (PLC), which up to now has been established in an heuristic way. {\color{reviewer2}The MPM model was initially proposed in the PLC context for outdoor channels in the band up to 20 MHz,} but its number of parameters becomes extremely large when used to model indoor channel frequency responses (CFR), which are much more frequency-selective than outdoor ones, and the band is extended to 80 MHz. This work proposes a fitting procedure that addresses this problem. It allows determining the model parameters that yield the best fit to each channel of a large database of single-input single-output (SISO) experimental measurements acquired  in typical home premises of different European countries. Then, the statistics of the MPM parameters are analyzed. The study unveils the relation between the model parameters and the main characteristics of the actual CFR like the frequency selectivity and the average attenuation. It also estimates the probability density function (PDF) of each parameter and proposes a fitting distribution for each of them. Moreover, the relationship among the main parameters of the model{\color{reviewer3}, as well as their impact on the performance of PLC communication systems} are also explored. Provided results can be helpful for the development of MPM-based models for indoor broadband PLC. 

\end{abstract}


\section{Introduction}
\IEEEPARstart{T}{he} increasing demand of high-speed multimedia services led to reconsider the indoor power delivery network as a data transmission medium. The \ac{PLC} technology is becoming an established solution since it exploits the existing wired infrastructure to convey high-speed data content, thus saving costs and deployment time. Furthermore, the latest standards and devices include advanced signal processing and enhanced transmission techniques, such as the exploitation of all the deployed network conductors and the bandwidth extension, that improve even further the achievable data rate, guaranteeing transmission rates in the order of gigabit per second at the physical layer \cite{Yonge13, G.9960,IEEE_P1901}. This high performance level motivates a renewed interest in developing novel and updated \ac{PLC} solutions, especially for home networking and smart grid applications \cite{Bush2013, Prasad2020, Righini2021, Canete2023,  Cortes23bis}. 

In order to support the growth of \ac{PLC} technology, it is fundamental to provide effective channel models, which are able to numerically emulate the properties of actual \ac{PLC} networks avoiding on field tests. Two basic approaches can be tackled in defining a model, i.e., bottom-up and top-down, and both of them can follow a deterministic or a statistical strategy \cite{Zimmermann02, Tonello07bis, Canete2011, Povedano23}. A deterministic strategy typically considers a specific network configuration, while a statistical strategy attempts to introduce some variability by means of statistical distributions. Concerning the possible approaches, the bottom-up method has a strong physical connection and describes the phenomena involved in the signal propagation by exploiting the transmission line theory. Typically, it provides a faithful channel representation, although it turns out to be a computationally onerous procedure. Some examples concerning the \ac{SISO} channel modeling can be found in \cite{Canete02,Galli06}, which propose a deterministic approach, and in \cite{Philipps99, Philipps00, Esmailian03, Tonello11, Canete2011}, which follow a statistical approach. The models discussed in \cite{Versolatto11} and \cite{Corchado16} extend the \ac{SISO} bottom-up strategy to the \ac{MIMO} case. 

Conversely, the top-down approach is an empirical method that fits an analytic function to a measurements database. This strategy facilitates the statistical extension by adopting a certain distribution for the parameters that describe the theoretical function. {\color{reviewer2}The \ac{MPM} is an example of this approach, and its validity in both outdoor and indoor \ac{PLC} scenarios is supported by the fact that power line cables are electrically long at the considered frequencies, resulting in a received signal that consists of multiples echoes caused by impedance mismatches at multiple points in the network. One of the first and more common \ac{MPM} \ac{SISO} models was proposed in \cite{Zimmermann02} and later statistically extended in \cite{Tonello12}.} Extensions of the \ac{SISO} formulation to the \ac{MIMO} case are presented in \cite{Canova2010} and \cite{Pagani16}. 

The \ac{PLC} channel response also exhibits a short-term variation synchronous with the mains voltage, which allows modelling it as a \ac{LPTV} system \cite{Canete06}. However, since the time variation is quite small compared to the duration of the impulse response, i.e., the channel response is underspread, the slow variation approximation can be made and the channel response can  be modeled as a series of time-invariant responses that repeat with the period of the mains signal. Hence, the \ac{MPM} model can be applied to each of these time-invariant responses. 


\subsection{Contribution}
{\color{reviewer2}The \ac{MPM} model has been widely used in \ac{PLC} scenarios \cite{Zimmermann02,OperaChannel05, Tonello07bis, Chien15, Huang2012, Gay2016}. Its validity for outdoor channels in the frequency band up to $20$ MHz was empirically assessed in \cite{Zimmermann02}, which provided the model parameters that yielded the best fit to a set of measured channels. Statistical extensions to indoor channels were made in \cite{Tonello07} and \cite{Tonello12}, which proposed channel generators that assumed certain \textit{a priori} probability distributions for the model parameters based on heuristic approaches. However, a rigorous measurement-based analysis neither of the \ac{MPM} capacity to model indoor scenarios in the band up to 80 MHz nor of the statistical distributions of the model parameters has been found in the literature.} 

In this context, we make three main contributions: 

\begin{itemize}
    \item We propose a method to determine the parameters of the \ac{MPM} model for indoor broadband \ac{PLC} channels in the band up to $80$ MHz. It addresses the problem of the extremely large number of parameters that results when the \ac{MPM} model in \cite{Zimmermann02} is extended to indoor scenarios and that are a consequence of the much more frequency selective behavior of indoor \acp{CFR} and of the bandwidth extension to $80$ MHz. 
    
    \item The proposed method has been used to perform a channel-by-channel fitting, determining the set of \ac{MPM} parameters that yields the best approximation to each of the channels of a large database consisting of 426 \acp{CFR} measured in indoor scenarios of different European countries.
            
    \item We give an in-depth analysis of the resulting \ac{MPM} parameters, showing their relation to the main channel characteristics like the frequency selectivity and the average attenuation, assessing their statistical behavior and the relation between them. Unlike the a posteriori parameters inference discussed in previous works, this study unveils the intrinsic properties of the \ac{MPM} parameter and can be helpful for the development of novel \ac{MPM} models or the refinement of existing ones. {\color{reviewer1} A discussion in this regard, and a study of the relationship between the main \ac{MPM} parameters and the performance of indoor \ac{PLC} systems are also provided.}
\end{itemize}


\subsection{Organization}
The paper is organized as follows. First, the \ac{MPM} is briefly recalled in Section \ref{sec:model}. Second, the assumptions and the method proposed to determine the \ac{MPM} parameters that give the best fit to each measured channel is detailed and justified in Section \ref{sec:proc}. The properties of the computed parameters, such as their relation to the channel characteristics, statistical behavior and relation between them are discussed in Section \ref{sec:statistical_analysis}. {\color{reviewer1}The application of the proposed results to a further development of \ac{MPM}-based channel generators and the analysis of the influence of the model parameters in the performance of \ac{PLC} systems is discussed in Section \ref{application_and_performance}}. Finally, conclusions follow in Section \ref{sec:con}.

\section{Background: The Multipath Propagation Model}
\label{sec:model}
The \ac{MPM} follows a top-down approach and describes the multipath propagation of the power signal into the \ac{PLC} medium that is due to the line discontinuities, such as branches, or unmatched loads. The analytic formulation of the \ac{CFR} can be expressed as \cite{Zimmermann02}
\begin{equation} 
  H\left(f\right) = A\sum_{i=0}^{N-1}{g_i}e^{-(a_0+a_1f^K)d_i} e^{-j2\pi f d_i/v},
 \label{eq:Hf}
\end{equation}
where $a_0$ and $a_1$ are the attenuation coefficients, $g_i$ and $d_i$ are the path gain and the path length of the $i$-th path\footnote{For the sake of notation simplicity in the subsequent fitting process, paths have been indexed as $0\leq i \leq N-1$, instead of $1\leq i \leq N$ as done in \cite{Zimmermann02}.}, respectively. The coefficient $g_i$ can be interpreted as the product of transmission and reflection coefficients by restricting $|g_i|\leq 1$ and by introducing the normalization coefficient $A>0$, which allows adjusting the attenuation of the resulting \ac{CFR} \cite{Zimmermann02}. The term $N$ denotes the number of paths and $K$ is the exponent of the attenuation factor related to the cable features. This work uses $N=2554$ (its rationale will be justified in Section \ref{sec:proc}.\ref{relation_among_L_M_N}), which is much larger than the values used in outdoor channels. Typical values of $K$ range between 0.5 and 1 \cite{Zimmermann02}, hence $K=1$ is assumed for the rest of this paper. The parameter  $\nu=c\epsilon_r$ is the propagation speed of light in the cables, where $c$ is the speed of light in the vacuum and $\epsilon_r$ is the relative dielectric constant of the insulator which envelops the conductors, which has been set to $\epsilon_r=1.5$ in order to take into account the non-uniformity of the dielectric \cite{Zimmermann02}.

{\color{reviewer3}Expression (\ref{eq:Hf}) is a truncated version of the actually infinite number of forward- and backward-traveling waves caused by the reflections due to impedance mismatch. The rationale for this truncation is that the amplitude of these waveforms decreases at each reflection. Hence, those that have experienced a larger number of reflections would have negligible amplitude and can be ignored. It must be emphasized that the path lengths, $d_i$, correspond to the actual distances traveled by the referred waves. For instance, in a simple network consisting of a voltage source and a mismatched transmission line of length $\mathcal{L}$ terminated in a load impedance, the distances travelled by the received waves are $d_i=(2i+1)\mathcal{L}$, with $i=0,1,2,\ldots$} \cite[Ch. 5]{Paul94}. 


\section{Method to Determine the \ac{MPM} Parameters}
\label{sec:proc}
In order to provide a statistical characterization of the parameters that describe the \ac{MPM} formulation in (\ref{eq:Hf}), the first step is to compute the \ac{MPM} parameters that give the best fitting of this analytic expression to each of the channels of the measurement database. {\color{reviewer1}It consists in a set of $426$ indoor \acp{CFR} corresponding to differential transmission and reception between the phase and neutral conductors (P-N), out of which $205$ have been acquired in Spain and $221$ where measured in the ETSI STF 410 MIMO PLC campaign described in \cite{Pagani16}.} Measured \acp{CFR} consists of $M=1262$ frequency samples given by $f_m = f_0 + m\Delta f$, with $f_0 = 1.0$ MHz, $0 \leq m\leq M-1$ and $\Delta f=62.5978$ kHz. 

{\color{reviewer1}
The characteristics of the measured channels are similar to those of the extensive campaign carried out in Italy \cite{Tonello14} and to the ones measured in the United States and reported in \cite{Galli11}. They also exhibit well-known properties of indoor \ac{PLC} channels, such as the lognormal distribution of the delay spread, which has a mathematical foundation \cite{Galli11}. The features of the employed channels are also coherent with the ones in \cite{Pagani16}. However, it is worth mentioning that while the $351$ channels used in the latter to derive a \ac{MIMO} channel model were also acquired in the referred ETSI campaign, they correspond to the P, N, E and CM (common mode) receiver ports described in \cite{ETSI_TR_101_562-3}, while in this work the conventional \ac{SISO} differential P-N links are employed.}

Since the model is described by many parameters, the following procedure is used to simplify the fitting strategy. First, the attenuation coefficients $a_0$ and $a_1$ are estimated. {\color{reviewer2}To this end, we consider an \ac{MPM} consisting on a single path. By setting $N=1$ in (\ref{eq:Hf}), it happens that $|A \cdot g_i|=1$, since $A$ is a normalization factor used to force $|g_i|\leq 1$. Furthermore, $N=1$ yields a linear attenuation profile from which the attenuation coefficients $a_0$ and $a_1$ can be estimated through the regression fit of each channel measurement. This linear attenuation profile can be assumed essentially determined by the longest path of the \ac{MPM}, which is denoted by $d_1=L$ (its value will be given in Section \ref{relation_among_L_M_N})}. The regression fit can be expressed as
\begin{equation}
	10\log_{10}|H(f)|^2 = -2(a_0+a_1f)L\cdot 10\log_{10}e = \alpha_0 + \alpha_1f,
	\label{eq:robust_fit}
\end{equation}
where $\alpha_0$ and $\alpha_1$ are the coefficients of the regression fit. Thus, the attenuation coefficients are computed as
\begin{equation}
	a_0 = -\frac{\alpha_0}{20 L\log_{10}e},\quad a_1 = -\frac{\alpha_1}{20 L\log_{10}e}.
	\label{eq:robust_fit_a}
\end{equation}

{\color{reviewer2} The presented procedure does not impose a rigid constraint on $a_0$, as there are multiple values of $a_0$, $g_i$ and $d_i$ that yield the same $g_i e^{-a_0 d_i}$. Hence, if the longest path of the considered channel is shorter than $L$, the small value of $a_0$  resulting from the employed procedure can be compensated with larger values of $g_i$. Regarding $a_1$, which is related to the slope of the attenuation profile, its value is generally small, since the attenuation in indoor \ac{PLC} channels does not exhibit the remarkable frequency dependent profile of outdoor channels \cite[Fig.5]{Cortes23}, as involved distances are much shorter and attenuation is mainly due to multipath propagation and not to the skin effect. The validity of these assumptions will be corroborated later by the fitting results.}

Second, the path lengths are discretized yielding uniformly spaced distances, 
\begin{equation}
	d_i = i\frac{L}{N} \quad \text{with } i=0, ..., N-1.
	\label{eq:d} 
\end{equation}
{\color{reviewer3}
It should be noted that actual path lengths are inherently discrete, as the lengths of the cable sections through which the waves propagate (forward and backward) are discrete quantities. Since indoor \ac{PLC} networks have complex topologies with numerous branches whose lengths are unknown \textit{a priori}, the uniform distribution appears to be the most appropriate choice. This assumption defines a grid of potential path lengths. Actual path lengths would be rounded to these grid elements. For the values of $L$ and $N$ used in this work (which will be given in Section \ref{relation_among_L_M_N}), the spacing is $1.25$ m. This means that the maximum rounding error of a path length would be $1.25/2=0.625$ m, which seems to be adequate even for small power line networks. Furthermore, it must be emphasized that the uniform grid is used as a starting point, those $d_i$ that do not correspond to actual (rounded) propagation distances within the considered network will have $g_i\approx0$ and will be pruned in subsequent steps of the procedure.}

Afterwards, the path gains, $g_i$, that give the best fitting to each measured channel have to be determined. This could be done by means of a \ac{LS} estimation that minimizes the \ac{RMSE} w.r.t. the measured channel in the set of $M$ measured frequencies $f_m$,
\begin{equation}
	\text{RMSE} = \sqrt{\frac{1}{M}\sum_{m=0}^{M-1}{|H(f_m)-\hat{H}(f_m)|^2}},
	\label{eq:RMSE}
\end{equation} where $H(f_m)$ and $\widehat{H}(f_m)$ represent the measured and fitted \ac{CFR} at frequency $f_m$, respectively. 

However, the \ac{RMSE} is dominated by the error in the largest values of the \ac{CFR}. Since the amplitude response of \ac{PLC} is strongly frequency-selective, this causes the fitting to be very accurate at lowly attenuated values to the detriment of the fitting of the highly attenuated values. To avoid this end and achieve a balanced error distribution along the entire amplitude range, we determine the path gains that minimize the \ac{NRMSE} defined as
\begin{equation}
	\text{NRMSE} = \sqrt{\frac{1}{M}\sum_{m=0}^{M-1}{\frac{|H(f_m)-\hat{H}(f_m)|^2}{|H(f_m)|^2}}}.
	\label{NRMSE}
\end{equation} 
The gains $g_i$ can be obtained by solving a \ac{WLS} problem formulated by expressing (\ref{eq:Hf}) in matrix form as follows,
\begin{equation}
	\underbrace{\begin{bmatrix}
			H(f_0) \\
			\vdots \\
			H(f_{M-1}) \\
			\conj{H}(f_{M-1}) \\
			\vdots \\
			\conj{H}(f_0) \\
	\end{bmatrix}}_{\mathbf{h}}
	=
	\underbrace{\begin{bmatrix}
			P_{0,0}  & \dots & P_{0,N-1} \\
			\vdots  &   & \vdots \\
			P_{M-1,0} & \dots  & P_{M-1,N-1} \\
			\conj{P}_{M-1,0}  & \dots & \conj{P}_{M-1,N-1} \\
			\vdots  &   & \vdots \\
			\conj{P}_{0,0} & \dots  & \conj{P}_{0,N-1} \\
	\end{bmatrix}}_{\mathbf{P}}
	\underbrace{\begin{bmatrix}
			g_0 \\
			\vdots \\
			g_{N-1}
	\end{bmatrix}}_{\mathbf{g}}
	\label{eq:Mat_def}
\end{equation}
where $P_{m,i} = \exp(-(a_0+a_1f_m)d_i)\cdot\exp(-j2\pi f_m d_i/v)$ and the scaling factor $A$ has been set to 1 by now. 

By defining the weighting matrix 
\begin{equation}
\begin{split}
\mathbf{W} = \text{diag}\bigg\{&\frac{1}{|H(f_0)|^2},\frac{1}{|H(f_1)|^2},\ldots,\frac{1}{|H(f_{M-1})|^2},\\
&\frac{1}{|H(f_{M-1}|^2}, \ldots,\frac{1}{|H(f_1)|^2},\frac{1}{|H(f_0)|^2}\bigg\},
\end{split}
\end{equation}
the values of $g_i$ can be obtained as
\begin{equation}
\mathbf{g} = \left(\mathbf{P}^{\HH}\mathbf{W}\mathbf{P}\right)^{-1}\mathbf{P}^{\HH}\mathbf{W}\mathbf{h}.
\label{eq:WLSg}
\end{equation}
 
\subsection{Dominant Paths Selection Procedure}
\label{decimation_process}
In order to fit each measured channel with a reduced set of parameters, paths with negligible contribution can be deleted, while keeping the most significant ones. To this end, a decimation procedure that iteratively discards the path with the smallest contribution to approximate the measured \ac{CFR} while keeping the \ac{NRMSE} below a certain threshold is employed. The index of the path to be discarded at each iteration, designated as $j$, is determined as 
\begin{equation}
	j = \argmin_{i}{ \left\{\sum_m |g_i| e^{-(a_0 + a_1f_m)d_i} \right\}}.
	\label{eq:criterion}
\end{equation}
{\color{reviewer2}The rationale for this criterion is that the contribution of a path to the overall \ac{CFR} is determined by $|g_i|e^{-(a_0 + a_1f)d_i}$ and not solely by the path length. Indeed, results presented in Section \ref{Statistical_Characterization} corroborate that the longest path is not generally the one with the smallest contribution.} 

The decimation procedure continues while the \ac{NRMSE} (dB) remains lower than a given threshold, which has been set to $-20$ dB. This ensures that the channel computed with a reduced set of paths still represents a very good approximation to the corresponding measured channel. Since the columns of the matrix $\mathbf{P}$ are not orthogonal, the path gains $g_i$ have to be recomputed after a path is discarded. To this end, the problem in (\ref{eq:Mat_def}) is reformulated by removing the $j$-th column of the matrix $\mathbf{P}$ and the new path gains are obtained. Table \ref{table:algorithm} summarizes the presented fitting and decimation procedure applied to each measured channel, where the expressions to obtain the values of $L$ and $N$ will be given in Section \ref{relation_among_L_M_N}.

When the decimation procedure ends, the path gains obtained for the considered measured channel are normalized by their maximum absolute value. This normalization factor provides the value for the parameter $A$. This way, the $g_i$ values, which represent the product of transmission and reflection coefficients, are in the range $g_i\in [-1,1]$, and the normalization coefficient $A\in(0,\infty)$. 

\begin{table}[h]
	\centering
	\renewcommand{\arraystretch}{1.5} 
	\caption{Algorithm of the fitting and path selection procedure applied to each measured channel}
	\begin{tabular}{l}
		\hline
		\textbf{Algorithm: MPM fitting procedure} \\
		\hline
		\textbf{Input:} $\mathbf{h}$ (measured \ac{CFR}) and $\text{NRMSE}_{\text{threshold}}=-20$ dB\\
            \textbf{Output:} $N$ (number of dominant paths), $\mathbf{d}$ (path lengths), $\mathbf{g}$ (path gains)\\ and $\mathbf{\widehat{h}}$ (fitted version of $\mathbf{h}$)\\
		\textbf{1:} Initialize $L$ and $N$ using (\ref{eq:L}) and (\ref{eq:Nbb}), respectively \\
            \textbf{2:} Initialize $\mathbf{d}$ using (\ref{eq:d}) \\
		\textbf{3:} Compute $a_0$ and $a_1$ using (\ref{eq:robust_fit_a}) \\				
		\textbf{4:} Compute $\mathbf{g}$ using (\ref{eq:WLSg}) \\
		\textbf{5:} Compute $\mathbf{\widehat{h}}=\mathbf{P}\mathbf{g}$\\
		\textbf{6:} Compute the \ac{NRMSE} using (\ref{NRMSE})\\		
		\textbf{7:} \textbf{while} \ac{NRMSE} (dB)$<\text{NRMSE}_{\text{threshold}}$ \textbf{do} \\
		\begin{tabular}{@{\hspace{1em}}l}
			\textbf{8:} Obtain $j$ using (\ref{eq:criterion})\\
			\textbf{9:} Calculate $\mathbf{P} = [\mathbf{P}(:,0:j-1),\mathbf{P}(:,j+1:N-1)]$ \\
			\textbf{10:} Update $\mathbf{g}$ using (\ref{eq:WLSg}) \\
			\textbf{11:} Update $\mathbf{\widehat{h}}=\mathbf{P}\mathbf{g}$ \\
			\textbf{12:} Compute the \ac{NRMSE} using (\ref{NRMSE})\\
			\textbf{13:} Update $N = N - 1$\\
		\end{tabular} \\
		\textbf{14:} \textbf{end while} \\
		\textbf{15:} Compute $\mathbf{P} = [\mathbf{P}(:,0:j-1),\mathbf{P}(:,j),\mathbf{P}(:,j+1:N-1)]$ \\
		\textbf{16:} Update $\mathbf{g}$ using (\ref{eq:WLSg})\\
            \textbf{17:} Update $\mathbf{\widehat{h}}=\mathbf{P}\mathbf{g}$ \\
		\hline
	\end{tabular}
	\label{table:algorithm}
\end{table}

{\color{reviewer1}
Table \ref{table:N_decimation} summarizes the results obtained when the aforementioned decimation path procedure is applied to fit the measured channels in the  frequency bands up to $20$ MHz and up to $80$ MHz. The former is included for comparison with the seminal work in \cite{Zimmermann02}, which applied the \ac{MPM} to outdoor scenarios in the frequency band up to $20$ MHz. The number of initial paths at the pre-decimation stage is $N=640$ for the $20$ MHz bandwidth and $N=2554$ for the $80$ MHz (as justified in the next subsection). As expected, the average number of paths after the decimation process is much smaller in the $20$ MHz case (less than half) than in the $80$ MHz one. Also, its variance  $\sigma_{N}$ is reduced. However, the number of paths is still much larger than in outdoor scenarios \cite{Zimmermann02}, where channels are less frequency selective because of the reduced number of derivations and increased attenuation (due to longer distances) that reduce the destructive effect of the multipath propagation. 

Regarding the efficiency of the decimation procedure, the number of paths is reduced to an average value (over the whole set of fitted channels) of $107.97$ in the $20$ MHz case and to $217.84$ in the $80$ MHz bandwidth. Interestingly, the former is about $16.8$\% of the initial $N$, while the latter is lower than $10$\% of the initial $N$. The higher efficiency in the $80$ MHz band is a consequence of the lower frequency selectivity of indoor \ac{PLC} channels above 30 MHz \cite{Cortes23bis}.}

\begin{figure}[!htb]
    \centering
    \includegraphics[width=\columnwidth]{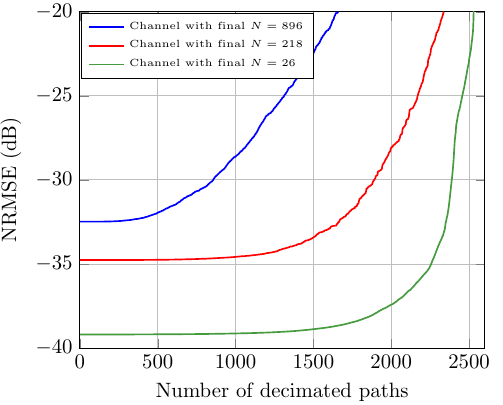} 
    \caption{{\color{reviewer3}Evolution of the \ac{NRMSE} with the number of decimated paths in three example channels. The delay spread values of the channels are $0.58$ $\mu$s for the one with $N=896$; $0.35$ $\mu$s for the one with $N=218$ and $0.11$ $\mu$s for the one with $N=26$.}}
    \label{fig:NRMSE_vs_iteration}
\end{figure}

{\color{reviewer3}To illustrate the increase in the \ac{NRMSE} with the number of decimated paths, Fig. \ref{fig:NRMSE_vs_iteration} shows the results obtained in three example channels\footnote{\color{reviewer3}To facilitate the reproducibility of these results, the measured channels and the \ac{MPM} parameters obtained in each iteration of the decimation process are provided in the GitHub repository https://github.com/franjavc/IEEE-TComm-2025-MPM-PLC.}: the one with the highest $N$ after the decimation process, another with a final value of $N$ similar to the average one shown in Table \ref{table:N_decimation} and the one with minimum value of $N$. Their delay spreads are $0.58$ $\mu$s, $0.35$ $\mu$s and $0.11$ $\mu$s, respectively. As observed, the \ac{NRMSE} increases approximately exponentially with the number of decimated paths. The lower the frequency selectivity of the channel, the lower the initial \ac{NRMSE}, the larger the flat region and the steeper the final increase of the \ac{NRMSE}. } 

{\color{reviewer3}The running time of the iterations of the decimation procedure\footnote{\color{reviewer3}Executed by MATLAB R2024b in a computer with Intel core i9-10900 CPU at 2.80 GHz and 32 GB of RAM.} quadratically decreases from approximately $7$ s for the first one to less than $0.1$ s for iterations above $2000$, since the size of the matrices in (\ref{eq:WLSg}) get smaller as the number of decimated paths increases. Nevertheless, it must be emphasized that the presented decimation process is not subject to any real-time constraint, as it might be the case of a real-time \ac{MPM} channel generator.
}

\begin{table}[t]
	\centering
	\caption{{\color{reviewer1} Summary of the dominant path selection procedure for channels up to $20$ MHz and $80$ MHz}}
{\color{reviewer1}    
	\begin{tabular}{ccc}
        \cmidrule[0.08em]{2-3}
        & \multicolumn{2}{c}{\textbf{Frequency band}}\\
            \toprule
		\textbf{Magnitude} & \textbf{20 MHz} & \textbf{80 MHz}\\
            \midrule
            Value of $N$ before decimation                       &640  &	2554\\
		\midrule
		$\overline{\textrm{NRMSE}}$ (dB) before decimation & -30.77 & -36.79\\
            \midrule
		$\sigma_{\textrm{NRMSE}}$ (dB) before decimation   & 7.69   &	7.99\\            
            \midrule
            Average $N$ after decimation                       &107.97  &	217.84\\
            \midrule
            $\sigma_{N}$ (dB) after decimation                 & 55.50	& 134.58\\				
            \midrule
            $\overline{\textrm{NRMSE}}$ (dB) after decimation  &-20.11  &	-20.05\\            
            \midrule
            $\sigma_{\textrm{NRMSE}}$ (dB) after decimation    & 0.1537 &	0.0687\\                        
		\bottomrule
	\end{tabular}  
}
	\label{table:N_decimation}
\end{table}

\begin{figure}[h!]
	\begin{subfloat}[]
		{\includegraphics[width=\columnwidth]{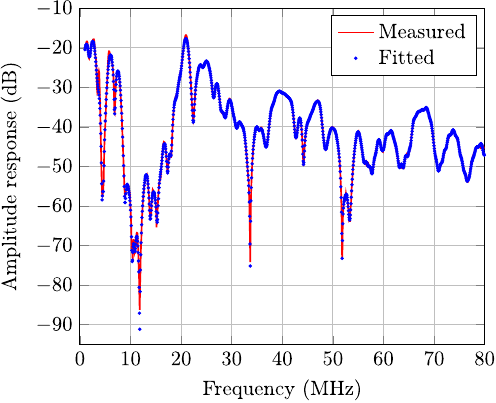}} 
		\label{fig:caso1_mod}
	\end{subfloat}%
	\begin{subfloat}[]
		{\includegraphics[width=\columnwidth]{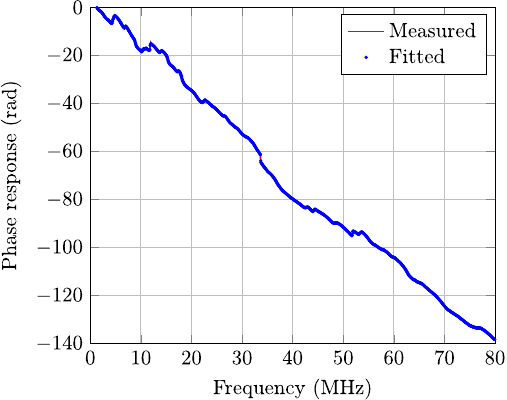}}
		\label{fig:caso1_phase}
	\end{subfloat}
\caption{Amplitude (a) and unwrapped phase (b) response of a representative measured channel and its corresponding fitting determined with $N=183$ dominant paths obtained with the proposed decimation procedure.}
\label{fig:case1}
\end{figure}

Presented results shows that the proposed decimation process achieves a good trade-off between modeling complexity and accuracy. For illustrative purposes, Fig. \ref{fig:case1} shows the amplitude and unwrapped phase of a measured \ac{CFR}, $H(f)$, and its corresponding fitted response, $\hat{H}(f)$, obtained with $N=183$ dominant paths determined using the proposed decimation process. As can be observed, the measured trace is almost veiled by the estimated one, which highlights the notable accuracy of the fitting process. 

The procedure described up to now assumed an initial number of paths, $N$, and a maximum path length, $L$. These parameters cannot be chosen arbitrarily and their selection criteria are described in the following section.

\subsection{Relation Among \textit{L}, \textit{M}, and \textit{N}}
\label{relation_among_L_M_N}
The maximum length $L$, the maximum number of paths $N$ depend on the considered frequency band and the frequency resolution of the measured \ac{CFR}. Hence, they are also related to the number of frequency samples, $M$, of the latter and must satisfy certain constraints which are now derived.

Since only a discrete and finite number of frequencies is available, the expression in (\ref{eq:Hf}) given for the continuous frequency domain can be rewritten for the discrete domain as
\begin{equation}
	H(f_m)=A\sum_{i=0}^{N-1}r_i e^{-j2\pi m\Delta fi\frac{L}{N\nu}},
	\label{eq:disc}
\end{equation}
where $\Delta f = (f_{M-1}-f_0)/(M-1)$ is the frequency resolution, while  $\tau=L/(N\nu)$ is the time resolution. Note that the real part of the exponential term has been embedded within the corresponding path gain in a unique variable named $r_i$. The adopted notation does not affect the analysis since the real exponential part acts only as an attenuation term that increases with frequency. This way, equation (\ref{eq:disc}) can be seen as a \ac{DFT} with frequency index $m$ and time index $i$. As known, the \ac{DFT} assumes the time signal belonging to the domain of integer multiples of $\tau$ and periodic of $N\tau$. Such a domain is denoted with $\mathbb{Z}(\tau)/\mathbb{Z}(N\tau)$. Hence, in the frequency domain
the signal is defined in $\mathbb{Z}(\Delta f)/\mathbb{Z}(N\Delta f)$, with $\Delta f = 1/(N\tau)$. Therefore, the relation between $L$ an the frequency samples $M$ is given by
\begin{equation}
	\Delta f=\frac{f_{M-1}-f_0}{M-1}=\frac{1}{N\frac{L}{N\nu}}\Rightarrow L=\frac{\nu}{\frac{f_{M-1}-f_0}{M-1}}=\frac{\nu}{\Delta f}.
 \label{eq:L}
\end{equation}
{\color{reviewer3}Expression (\ref{eq:L}) can be also derived from physical considerations. The time it takes for an input impulse to the channel to reach the receiver following the longest path (i.e., the one with the largest number of reflections) is $\tau=L/\nu$. The maximum duration of the channel impulse response that can be obtained from a sampled version of the \ac{CFR} at multiples of $\Delta f$ while avoiding time-domain aliasing is $\tau=1/\Delta f$. Equating both expressions yields $L=\nu/\Delta f$.}

For $f_{M-1}=79.9358$ MHz and $f_{0}=1$ MHz, $M=1262$ and the propagation speed is \mbox{$\nu= 2\cdot 10^8$m/s}, the maximum possible length that can be assumed is $L=3195$ m. {\color{reviewer3}Since physical distances in indoor \ac{PLC} networks are on the order of tens or few hundreds of meters, the limit of over $3$ kilometers is large enough to account for the longest distance traveled by all forward- and backward-propagating waves received with significant power.}

Referring to the \ac{DFT} representation of the \ac{CFR} reported in (\ref{eq:disc}), the \ac{DFT} frequency period, i.e., $N\Delta f$, must be greater or equal than two times the maximum channel frequency, i.e., $f_{M-1}$, in order to avoid aliasing. {\color{reviewer2}Thus, the relationship among $N$ and $L$ turns out to be
\begin{equation}
	N\Delta f=\frac{1}{\tau}\ge 2 f_{M-1} \Rightarrow N\ge \frac{2 f_{M-1}L}{\nu}=\frac{f_{M-1}}{\Delta f}.
	\label{eq:Nb}
\end{equation}}
As seen, $N$ is determined by the considered bandwidth and the frequency resolution of the measurements. If the aforesaid values for $f_{M-1}$ and $\nu$ are considered, the relation in (\ref{eq:Nb}) becomes $N\ge0.8L$.
A final relation among $N$ and $L$ can be noticed by rewriting the imaginary part of the exponential term in (\ref{eq:Hf}) as
\begin{equation}
	e^{-j2\pi fd_i/\nu}=e^{-j2\pi kd_i/\lambda}
\end{equation}
where $\lambda=\nu /\Delta f$ is the maximum possible wavelength, which is related to the frequency resolution $\Delta f$. The minimum
observable wavelength is $\lambda_{min}=\nu /f_{M-1} $. Hence, to avoid numerical errors, the shortest (non-zero) path length $d_1=\frac{L}{N}$ must be greater
than $\lambda_{min}/2$, according to the sampling theorem. This provides the last relationship, expressed by
\begin{equation}
	d_1=\frac{L}{N}\ge\frac{\lambda_{min}}{2} \Rightarrow N\le \frac{2 f_{M-1}L}{\nu}=\frac{f_{M-1}}{\Delta f}.
	\label{eq:Nbb}
\end{equation}
{\color{reviewer2}Expression (\ref{eq:Nbb}) can be also derived from physical considerations. Assuming uniform distances, the time difference between the closest received paths is $\frac{L/N}{\nu}$. In order for these paths to be resolvable, this time must be greater or equal than the minimum sampling period, $T_s=\frac{1}{2f_{M-1}}$. This yields $\frac{L/N}{\nu}\geq \frac{1}{2f_{M-1}}$, which equals (\ref{eq:Nbb}).}

Given the specifications of the measured \ac{CFR}, as well as the relations (\ref{eq:Nb}) and (\ref{eq:Nbb}), the relationship among
$L$ and $N$ must hold with equality, i.e., $N=0.8 L=2554$ for the considered database.

\section{Statistical Analysis of the \ac{MPM} Parameters}
\label{sec:statistical_analysis}
This section analyzes the statistical properties of the \ac{MPM} parameters obtained according to the fitting procedure detailed in Section \ref{sec:proc} for the overall set of measured channels. First, the relation of some model parameters to the main quantities used to characterize the channel response, such as the delay spread and the average channel gain. Then, the \ac{PDF}
of the \ac{MPM} parameters is estimated and, finally, the statistical relation between the model parameters is studied. 

\subsection{Relation to the Channel Characteristics}
\begin{figure*}[htb]
	\centering
	\begin{subfloat}[]
		{\includegraphics[width=0.299\textwidth]{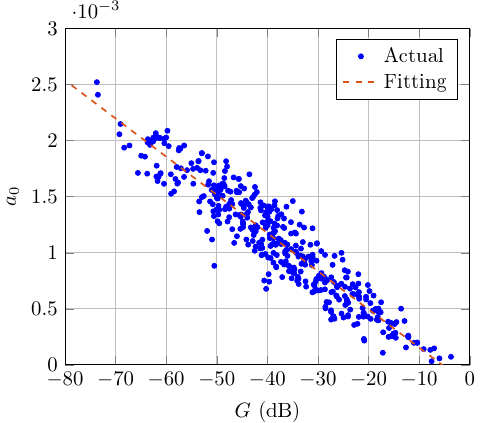}}
	\end{subfloat}
	\hfill
	\begin{subfloat}[]
		{\includegraphics[width=0.39\textwidth]{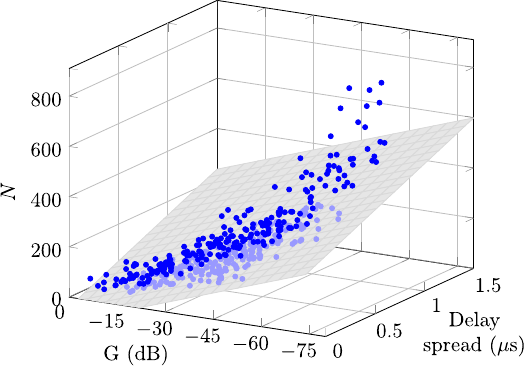}}
	\end{subfloat}
        \hfill
        \begin{subfloat}[]
		{\includegraphics[width=0.299\textwidth]{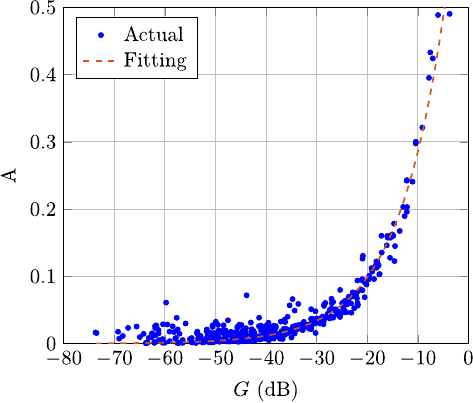}}
	\end{subfloat}
	\caption{Scatter plot of (a) the value of $a_0$ vs the average channel gain of each measured channel, $G$ (dB); (b) the number of dominant paths vs the average gain and the delay spread of the channels; (c) value of $A$ vs the average channel gain of each measured channel.}
	\label{fig:a0_N_A}
\end{figure*}

The parameters $a_0$ and $a_1$ were obtained from the regression fit of the amplitude response of the measured channels, with $a_1$ being related to the slope of the attenuation profile and $a_0$ with the y-intercept. Hence, the latter parameter has to be somehow related to the average channel gain, defined as ${G\textrm{ (dB)}=\frac{1}{M}\sum_{m=0}^{M-1}20\log_{10}{|H(f_m)|}}$. This relation can be clearly observed in Fig. \ref{fig:a0_N_A} (a), where the scatter plot of $a_0$ vs ${G\;\textrm{(dB)}}$ is depicted, along with its robust fit regression line, whose expression is given in Table \ref{table:fittings}. As expected, channels with larger average gains need lower values of $a_0$ to be synthesized. 

The number of dominant paths needed to fit a given \ac{CFR} is related to the frequency selectivity of the channel, e.g., a single path suffices to synthesize a \textit{flat} channel. The frequency selectivity of a channel is usually quantified by means of the coherence bandwidth or its inversely related magnitude, the delay spread (also referred to as multipath spread) \cite{Molish2011}. In \ac{PLC} these magnitudes are known to be related to the average channel gain \cite{Cortes23bis}: channels with lower average channel gains tend to have larger delay spreads. Hence, these magnitudes are expected to be related to the number of dominant paths, $N$. Fig. \ref{fig:a0_N_A} (b) illustrates  this relation by showing the scatter plot of the number of dominant paths vs the average channel gain and the delay spread of the corresponding measured channel. Its robust regression plane, whose expression is given in Table \ref{table:fittings}, is also depicted for reference. The goodness of this fit is confirmed by the \ac{NRMSE} between the actual and predicted values of $N$, which is $-22$ dB. 


The value of the normalization parameter $A$ is also related to the average channel gain, as shown in Fig. \ref{fig:a0_N_A} (c), which depicts the scatter plot of $A$ vs $G$ (dB) and its corresponding exponential fitting function. The parameters of the latter, which have been obtained by means of a \ac{LS} fitting, are given in Table \ref{table:fittings}. As observed, the lower the average channel gain, the lower the value of $A$ needed to synthesize the channel. 


\newcolumntype{C}[1]{>{\centering\arraybackslash}m{#1}}
\begin{table*}[b]
	\centering
	\caption{Parameters of the fitting expressions displayed in Fig. \ref{fig:a0_N_A}.}
	\begin{tabular}{c C{2.5cm} C{2.5cm} C{2.5cm}}
		\toprule
		\textbf{Fitting expression} & $\bm{\alpha}$ & $\bm{\beta}$ & $\bm{\gamma}$\\
		\midrule
		$\widehat{a}_0=\alpha + \beta \cdot G\textrm{ (dB)}$ & $-1.8669\cdot10^{-4}$ & $-3.4066\cdot 10^{-5}$ & -\\
		\midrule
		$\widehat{N}=\alpha +  \beta \cdot ds\textrm{ ($\mu$s)} + \gamma \cdot G\textrm{ (dB)}$ & $40.4009$ & $185.7535$ & $4.5097$\\
		\midrule
		$\widehat{A}=\alpha \cdot e^{\beta \cdot G\textrm{ (dB)}}$ & $8.2517 \cdot 10^{-1}$ & $1.0636 \cdot 10^{-1}$ & - \\
        \bottomrule
	\end{tabular}
	\label{table:fittings}
\end{table*}

\subsection{Statistical Characterization}
\label{Statistical_Characterization}
This subsection estimates the main moments and the \ac{PDF} of the \ac{MPM} parameters, computed for the overall set of measured channels, that have been obtained using the fitting procedure presented in Section \ref{sec:proc}. For each empirical (estimated) \ac{PDF}, the probability distribution that gives the best fit is given. To this end, the most common distribution functions are considered: beta, Birnbaum-Saunders, exponential, gamma, \ac{GEV}, Gumbel, inverse Gaussian, logistic, log-logistic, lognormal, Nakagami, normal, Poisson, Rayleigh, Rician, $t$ location-scale and Weibull. For each distribution, the parameters that yield the best fit to the set of \ac{MPM} parameters are obtained by means of the \ac{ML} estimation assuming independent samples \cite{Bickel2015}. Afterwards, the Anderson-Darling goodness-of-fit test is applied and the distribution yielding the highest $p$-value is selected. When the test gives the same $p$-value for all distributions or the null hypothesis is rejected for all of them, the one with the highest value of the likelihood function is selected. For some parameters, more elaborate models consisting of mixture distributions have been considered. 

The mean and standard deviation of the \ac{MPM} parameters are given in Table \ref{table:m&s}. The fitting statistical distributions with their corresponding parameter values are summarized in Table \ref{table:param}. These reported results allow future replicability of this study and simplify the development of improved statistical models.

\begin{table}[t]
	\centering
	\caption{Mean and standard deviation of the MPM parameters}
	\begin{tabular}{ccc}
		\toprule
		\textbf{MPM parameter} & \textbf{mean} & \textbf{standard deviation}\\
		\midrule
		Atten. coeff. ($a_0$) & 0.0011 & 5.0089e-4 \\
		\midrule
		Atten. coeff. ($a_1$) & 6.0953e-12 & 4.1137e-12\\		
		\midrule
		Norm. coeff. ($A$) & 0.0422 & 0.0649 \\		
		\midrule
		Number of dominant paths ($N$) & 217.8427 & 1.3457 \\	
		\midrule
		Path gains ($|g_i|$) & 0.1384 & 0.1859 \\	
		\midrule
		Path lengths ($d_i$) & 343.9491 & 633.4392 \\	
		\bottomrule
	\end{tabular}
	\label{table:m&s}
\end{table}

\begin{figure*}[ht]
	\centering
	\begin{subfloat}[]
		{\includegraphics[width=0.32\textwidth]{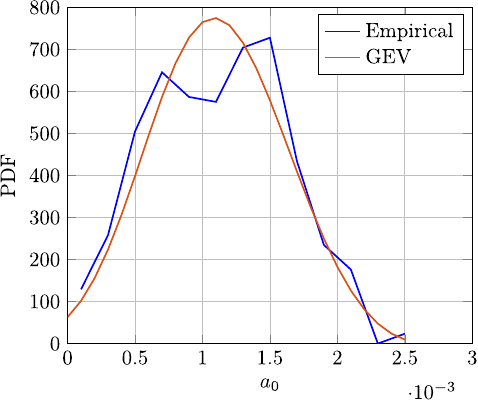}}
	\end{subfloat}
	\hfill
	\begin{subfloat}[]
		{\includegraphics[width=0.32\textwidth]{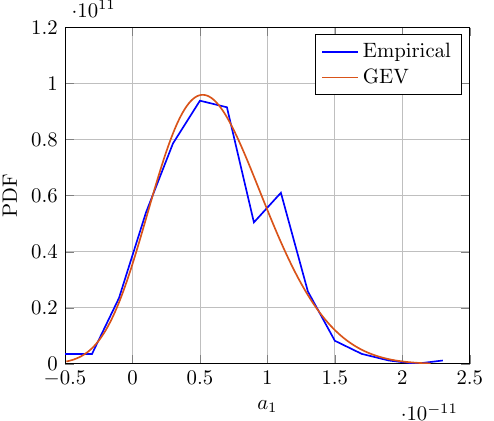}}
	\end{subfloat}
        \hfill
        \begin{subfloat}[]
		{\includegraphics[width=0.32\textwidth]{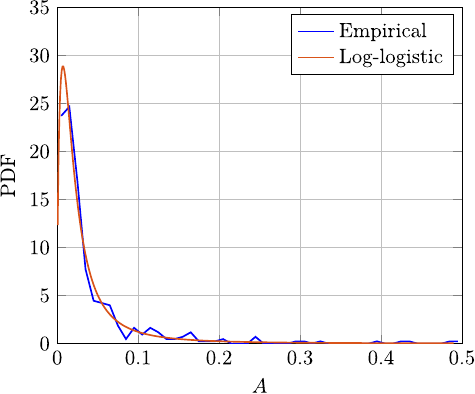}}
	\end{subfloat}
	\caption{Empirical and fitted \acp{PDF} of the attenuation coefficients $a_0$ and $a_1$ and the normalization coefficient $A$.}
	\label{fig:hist1}
\end{figure*}

\begin{figure*}[ht]
	\centering
	\begin{subfloat}[]
		{\includegraphics[width=0.343\textwidth]{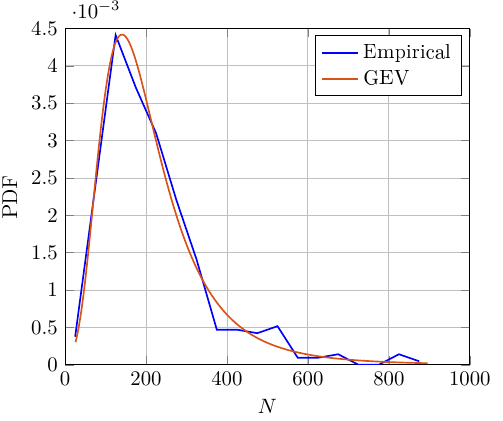}}
	\end{subfloat}        
        \hfill
	\begin{subfloat}[]
		{\includegraphics[width=0.31\textwidth]{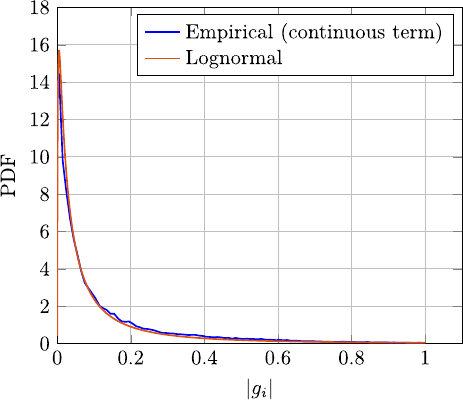}}
	\end{subfloat}
	\hfill
	\begin{subfloat}[]
		{\includegraphics[width=0.317\textwidth]{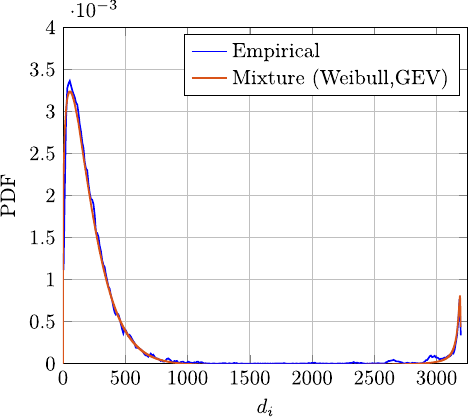}}
	\end{subfloat}
	\caption{Empirical and fitted \acp{PDF} of the number of dominant paths, $N$, the modulus of the path gains $|g_i|$ and the path lengths $d_i$.}
 	\label{fig:hist2}
\end{figure*}

The \acp{EPDF} of the \ac{MPM} parameters are reported in Fig. \ref{fig:hist1} and Fig. \ref{fig:hist2}. Regarding $a_0$ and $a_1$, whose \acp{EPDF} are show in Fig.  \ref{fig:hist1} (a) and (b), respectively, the distribution that gives the best fit is the \ac{GEV} ($p$-value$=0.286$ for $a_0$ and 0.8941 for $a_1$), although the normal distribution also gives a high $p$-value (0.1967 for $a_0$ and 0.5664 for $a_1$). In practice, the \ac{GEV} is a family of three types of extreme value distributions, namely type I, or Gumbel (or log-Weibull) \cite{Gumbel1954}, type II, or Fr\'echet, and type III, or Weibull distribution. This result makes sense since the \ac{GEV} is the limit distribution of properly normalized maxima of a sequence of independent and identically distributed random variables. The \ac{GEV} distribution for $a_0$ is justified since it represents the maximum (or the minimum in case of negative values of $a_1$) of the channel attenuation profile for the identified trend of each measured \ac{CFR}. 

Regarding the normalization coefficient, $A$, whose \ac{EPDF} is displayed in Fig. \ref{fig:hist1} (c), the Log-logistic gives the largest $p$-value (0.909) although, although the \ac{GEV} and the lognormal also give high values ($p$-values $0.788$ and $0.4536$, respectively). The latter is not surprising, as $A$ is determined as the maximum of the absolute value of the path gains, $|g_i|$, of each fitted channel. 

Regarding the probability distribution of the number of dominant paths, $N$, while this magnitude can only take non-negative integer values, discrete distributions like the binomial and Poisson ones give very poor fitting to the actual set of values. Hence, $N$ has been considered as a discretized version of a continuous random variable. This same approach has been also applied to fit the path length, $d_i$. By doing so, the \ac{GEV} gives the best fit ($p$-value$=0.86752$) to the number of dominant paths, $N$, whose \ac{EPDF} is displayed in Fig. \ref{fig:hist2} (a). As a matter of fact, $N$ represents the maximum number of dominant paths that actively contribute to reconstruct each measured \ac{CFR}. 

Concerning the characterization of the path gain, its absolute value, $|g_i|$, is considered for the sake of simplicity, as the \ac{EPDF} of $g_i$ shows a symmetric behavior around the $y$-axis. The statistical characterization of $|g_i|$ must take into account that the normalization process used to force $g_i\in [-1,1]$ in each fitted channel causes the probability of $|g_i|=1$ to be non-zero. Hence, $|g_i|$ has to be modeled by a mixed \ac{RV} consisting of a continuous component and a discrete term at $|g_i|=1$. Fig. \ref{fig:hist2} (b) shows that the continuous term of the \ac{EPDF} can be fitted by a lognormal \ac{RV}. Accordingly, the overall \ac{PDF} can be expressed as $f(|g_i|)=\pi_0 q(|g_i|;\mu,\sigma)+\pi_1 \delta(|g_i|-1)$, where $q(\cdot;\mu,\sigma)$ denotes the \ac{PDF} of a lognormal \ac{RV} with parameters $\mu$ and $\sigma$ and $\delta(\cdot)$ is the Dirac delta function. The value of these parameters are given in Table \ref{table:param}. 

The \ac{EPDF} of $d_i$, displayed in Fig. \ref{fig:hist2} (c),  exhibits a bimodal behavior, with maxima around $d_i=50$ m and $d_i=3195$ m. The latter is an effect of the use of the discrete frequency domain version of (\ref{eq:Hf}), which obliges to set a limit for the largest path length. Since the observed bimodal behavior cannot be obtained with any of the distributions indicated above, we have resorted to mixture models. It has been found that the Weibull distribution gives a good fitting to the values of $d_i\leq1500$ while the \ac{GEV}, shifted and reflected about $d_{N-1}=\frac{N-1}{N}L$, adequately fits the values of $d_i>1500$. Hence, the \ac{PDF} of $d_i$ can be modeled as $f(d_i)=\pi_0 g_0(d_i;\lambda_0, k_0) + \pi_1 g_1(d_{N-1}-d_i;k_1,\sigma_1,\mu_1)$, where $g_0(\cdot;\lambda_0, k_0)$ denotes the \ac{PDF} of a Weibull distribution with scale and shape parameters $\lambda_0$ and $k_0$, respectively, and $g_1(\cdot;k_1,\sigma_1,\mu_1)$ is the \ac{PDF} of a \ac{GEV} distribution with location, scale and shape parameters $k_1$, $\sigma_1$ and $\mu_1$, respectively, whose values are given in Table \ref{table:param}. As shown in Fig. \ref{fig:hist2} (c), the \ac{PDF} of the proposed mixture gives a good fitting to the empirical one.

\begin{figure*}[t]
	\begin{subfloat}[]
		{\includegraphics[width=0.33\textwidth]{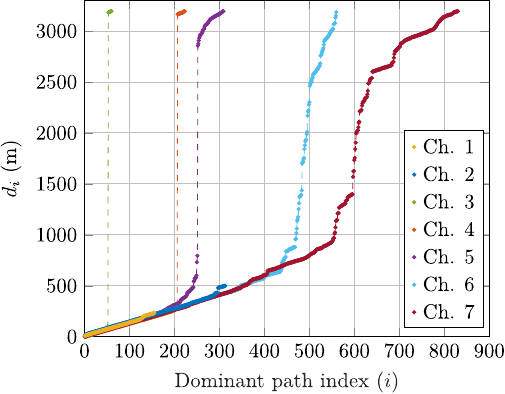}} 
	\end{subfloat}%
	\begin{subfloat}[]
		{\includegraphics[width=0.315\textwidth]{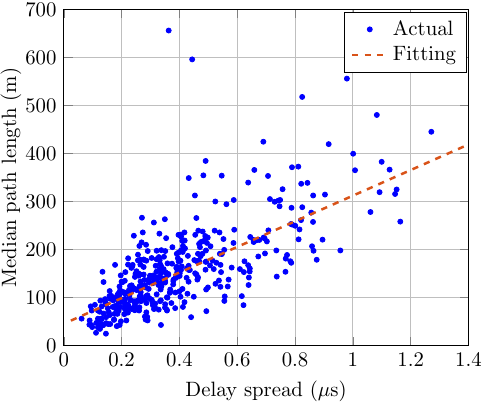}}
	\end{subfloat}
        \begin{subfloat}[]
		{\includegraphics[width=0.315\textwidth]{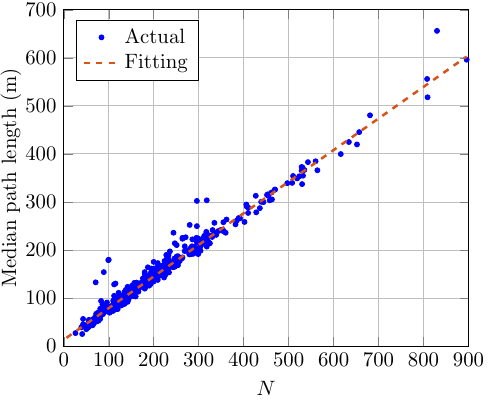}}
	\end{subfloat}
\caption{(a) Dominant path length vs dominant path index for some illustrative measured channels; (b) scatter plot of the median value of the dominant path length vs the delay spread for the set of measured channels; (c) scatter plot of the median value of the dominant path length vs the number of dominant paths for the set of measured channels.}
\label{fig:path_length_analysis}
\end{figure*}

\begin{table}[th]
        \setlength{\tabcolsep}{4.5pt}
	\centering
	\caption{Fitting distributions of the \ac{MPM} parameters}
	\begin{tabular}{ccc}
		\toprule
		\begin{tabular}{c}\textbf{MPM}\\ \textbf{parameter}\end{tabular} & \textbf{Distribution} & \begin{tabular}{c}\textbf{Distribution}\\ \textbf{parameters}\end{tabular}\\
		\midrule
		\begin{tabular}{c}Atten. coeff.\\ $a_0$\end{tabular} & GEV & 
		\begin{tabular}{c} 
			$k=-0.2744$\\ 
			$\sigma=4.9492\cdot 10^{-4}$\\
			$\mu=9.3370\cdot 10^{-4}$ 
		\end{tabular}\\
		\midrule
		\begin{tabular}{c}Atten. coeff.\\$a_1$\end{tabular} & GEV & 
		\begin{tabular}{c} 
			$k= -0.1781$\\ 
			$\sigma=3.8980\cdot 10^{-12}$\\
			$\mu=4.4536\cdot 10^{-12}$ 
		\end{tabular}\\		
		\midrule
		\begin{tabular}{c}Norm. coeff.\\ $A$\end{tabular} & Log-logistic & 
		\begin{tabular}{c} 
			$\mu=-3.8584$\\
			$\sigma=0.6710$ 
		\end{tabular}\\		
		\midrule
		\begin{tabular}{c}Number of \\dominant paths\\ $N$\end{tabular} & GEV & 
		\begin{tabular}{c} 
		$k=0.1641$\\ 
		$\sigma=84.3237$\\
		$\mu=153.5548$ 
		\end{tabular}\\	
		\midrule				
            \begin{tabular}{c}Path gains\\$|g_i|$\end{tabular} & \begin{tabular}{c}Mixture\\ (lognormal, $|g_i|=1$)\end{tabular}& 
		\begin{tabular}{c}
                $\pi_0=1-\pi_1$\\
                $\pi_1=4.6017\cdot10^{-3}$\\
			$\mu=-2.9139$\\
			$\sigma=1.5445$
		\end{tabular}\\	
            \midrule
            \begin{tabular}{c}Path lengths\\ $d_i$\end{tabular} & \begin{tabular}{c}Mixture\\ (Weibull, GEV)\end{tabular} & 
		\begin{tabular}{c} 
                $\pi_0= 9.5074 \cdot 10^{-1}$\\
                $\pi_1= 1-\pi_0$\\
                $\lambda_0=2.1894 \cdot 10^2$\\
                $k_0=1.2314$\\                                
			$k_1=1.3432$\\ 
		      $\sigma_1=42.5933$\\
		      $\mu_1=27.4299$ 
		\end{tabular}\\	            
		\bottomrule
	\end{tabular}
	\label{table:param}
\end{table}

In order to get an in-depth analysis of the behavior of $d_i$, Fig. \ref{fig:path_length_analysis} (a) displays the path length, $d_i$, vs the dominant path index, $i$, for some illustrative measured channels. For the sake of clarity, the markers corresponding the same channel have been linked by a dashed line of the the same color. As seen, all channels have a set of paths whose length is uniformly distributed from zero (or almost zero) and a given value that depends on the channel. However, some channels, like Ch. 3-7, also have a set of paths with very large length. In many cases, like Ch. 3 and Ch. 4, there are no paths with intermediate distances. This behavior is related to the frequency selectivity of the channel: the larger the frequency selectivity, the larger the distances of the paths required to synthesize is. This can be observed in Fig. \ref{fig:path_length_analysis} (b), where the scatter plot of the median value of the dominant path lengths vs the delay spread of each measured channel is given. As observed, channels with higher delay spread values (larger frequency selectivity) need larger paths to be synthesized. Furthermore, it is also expected that the higher the number of dominant paths, the larger the median length of the paths. Fig. \ref{fig:path_length_analysis} (c) confirms this end by showing a strong linear relationship between them.

\subsection{Comparison with Other Works}
The \ac{MPM} parameter distributions that have been just presented are herein compared with those heuristically or empirically obtained in previous scientific works. For example the presented statistical analysis confirms the assumptions made in \cite{Tonello12}, which statistically extended the \ac{MPM} in (\ref{eq:Hf}) relying on the initial idea presented in \cite{Tonello07bis}. Hence, it was inferred in \cite{Tonello12} that the path gains $g_i$ turn out to be lognormally distributed variables multiplied by a random flip sign. The main difference is found for the statistics of $N$, supposed to have a Poisson distribution in \cite{Tonello12}, but that results into a \ac{GEV} one. This difference was already hypothesized in \cite{Tonello12} and is also justified by the different considered databases. The path lengths, $d_i$, were assumed to be lognormally distributed in \cite{Tonello12}, while the mixture of a Weibull and \ac{GEV} has been obtained in this work. Additionally, the characterization method proposed in this paper provides a statistics for the other parameters, i.e., $a_0$, $a_1$, and $A$. 

Other results concerning the \ac{MPM} parameters distribution are listed in \cite[Table 1]{Hashmat2011}, that extends the model in (\ref{eq:Hf}) to the \ac{MIMO} transmission. In this case, the path gains $g_i$ are assumed uniformly distributed in $[-1, 1]$ and the attenuation coefficient $a_1$ turns out to be constant. While, $K$ is normally distributed and $a_0$ exhibits a shifted exponential distribution. Instead, the channel median $A$ is uniformly or exponentially distributed depending on the considered channels. The same results are also summarized in \cite[Table 2.10]{Lampe16}.

\subsection{Relationship Between Parameters}
\label{relation_between_parameters}
The development of \ac{MPM}-based models requires a complete characterization of the involved parameters. This includes both their marginal \acp{PDF}, which have given in the previous subsection, and their conditional distribution, which are examined in this one. 

\begin{figure}[!htb]
    \centering
    \includegraphics[width=\columnwidth]{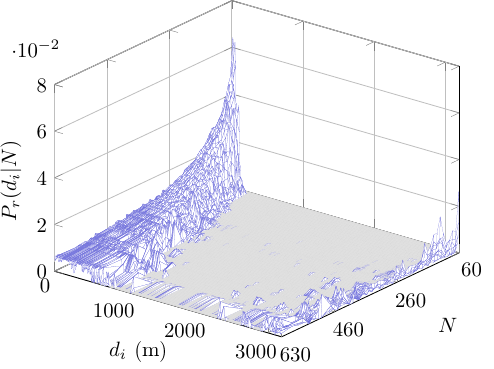} 
    \caption{Conditional probability of the path length, $d_i$, given the number of dominant paths, $N$, estimated over the set of measured channels.}
    \label{fig:3D_di_vs_N_ds}
\end{figure}

First, the relation between the path length, $d_i$, and the number of dominant paths, $N$, is explored. To this end, the probability distribution of the path length conditioned on $N$, $f\left(d_i|N\right)$ is estimated. Obtained results are given in Fig. \ref{fig:3D_di_vs_N_ds}, which gives additional details on the already observed bimodal behavior of the \ac{EPDF} of $d_i$. As seen, there are almost no paths with lengths in the range from $1000-3000$ m, approximately, except for the case where $N$ is very large, in which some paths with lengths in this range can be found. The values of $d_i$ are concentrated in the low length region and in the region around 3000 m, being (approximately) uniformly distributed from zero (or almost zero) in the low length region. Furthermore, the range of path lengths in the low region increases with the number of dominant paths. 

\begin{figure}[!htb]
    \centering
    \includegraphics[width=\columnwidth]{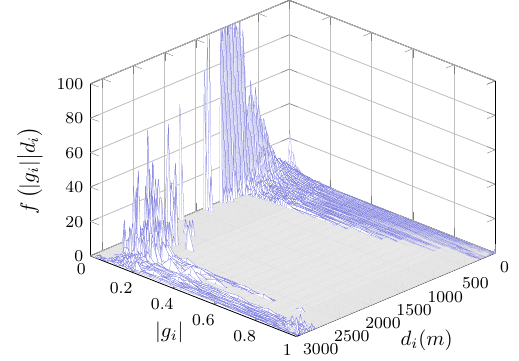} 
    \caption{Conditional \ac{PDF} of $|g_i|$ given the path length, $d_i$, estimated over the set of measured channels.}
    \label{fig:3D_gi_vs_d_i_ds}
\end{figure}

Next, the relation between the absolute value of the path gain and the path length is assessed by estimating the conditional \ac{PDF} of $|g_i|$ given $d_i$, $f\left(|g_i|\big| d_i\right)$. Results are displayed in Fig. \ref{fig:3D_gi_vs_d_i_ds}. Interestingly, $|g_i|$ tends to a uniform distribution in the interval $(0,1)$ for very low  values of $d_i$. As the path length increases, the range of values of $|g_i|$ decreases until it tends, again, to a uniform one in the interval $(0,1)$ for path lengths around $3000$ m. There is a clearly noticeable gap for path lengths between $1000-2500$ (approximately), where the few existing paths have very low gains. 

Finally, the correlation between the path gains is explored. Since the decimation process described in Section \ref{decimation_process} yields a different number of paths for each measured channel, the following procedure is employed to allow comparing results from different channels. In order to distinguish the \ac{MPM} model parameters before the decimation process is applied from the ones resulting after it, let us denote the latter by using tilded symbols. Hence, the original set of path lengths and path gains are expressed as $d_n$ and $g_n$, respectively, with $0\leq n \leq N-1$, where $N=$2554, and their corresponding sets after the decimation process as $\tilde{d}_i$ and $\tilde{g}_i$, respectively, with $0\leq i \leq \tilde{N}$, where $\tilde{N}$ is the number of dominant paths after the decimation process. These magnitudes are related as
\begin{equation}
    g_n=
    \begin{cases}\tilde{g}_i & \text{if }  \exists i \;|\; \tilde{d}_i=d_n\\ 
                           0 & \text{otherwise}
    \end{cases}.
\end{equation}

\begin{figure}[!htb]
    \centering
    \includegraphics[width=\columnwidth]{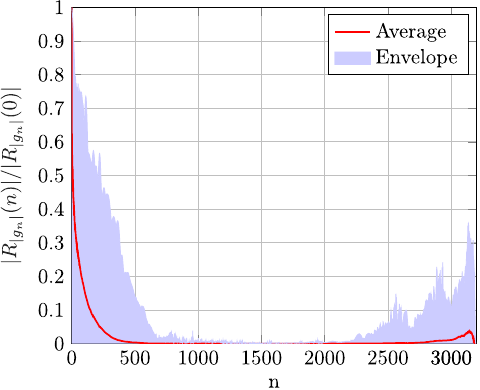} 
    \caption{Average and envelope of the normalized deterministic \ac{ACF} of $|g_n|$ computed over the set measured channels.}
    \label{fig:R_gi_vs_di}
\end{figure}

In order to explore the correlation among the path gains, the deterministic \ac{ACF} of $|g_n|$ for each measured channel, $R_{|g_n|}(n)=\frac{1}{N}\sum_{m=0}^{N-n-1}|g_{n+m}||g_{m}|$, for $0\leq n \leq N$, is computed and normalized. Fig. \ref{fig:R_gi_vs_di} depicts the its average value (over the set of measured channels) and the envelope of the \ac{ACF} of all channels. As seen, the \ac{ACF} decreases very rapidly, with values below 0.5 for $n\geq8$ and below 0.3 for $n\geq40$, and slightly regrows for large values of $n$ due to the behavior already shown in Fig. \ref{fig:3D_gi_vs_d_i_ds}. 

\section{Applicability of the study and influence on the performance of \ac{PLC} systems}
\label{application_and_performance}
 This section discusses the use of the statistical analysis of the \ac{MPM} parameters given in this work to the development of a stochastic \ac{MPM}-based channel generator and also explores the relation between the model parameters and  the performance of indoor \ac{PLC} systems. 

{\color{reviewer1}
\subsection{Application to the development of a stochastic \ac{MPM}}
\label{application}
While the development of a statistical \ac{MPM}-based channel generator is beyond the scope of this work, information provided herein can be used as a starting point to generate realizations of \acp{CFR} according to the \ac{MPM}. To this end, the steps given below can be followed:  

\begin{enumerate}
\item Generate a realistic value of the average channel gain, ${G\;\textrm{(dB)}}$. To this end, the \ac{EPDF} of ${G\;\textrm{(dB)}}$ for the set of measured channels is given in Fig. \ref{fig:mesured_channels_charac} (a), along with the \ac{PDF} of the distribution that gives the best fit, whose parameters are provided in Table \ref{table:G_ds_measured}. 

\item Generate a realistic value of the delay spread. This can be obtained by leveraging the relation between the average channel gain and the delay spread depicted in Fig. \ref{fig:mesured_channels_charac} (b), where the logarithm is taken to avoid the heteroscedasticity in the regression line, which is given by 
\begin{equation}
	\ln(\widehat{\textrm{ds}} (\mu \textrm{s}))=\alpha  + \beta \cdot G (\textrm{dB}),
	\label{eq:fit_log_ds_G}
\end{equation}
with $\alpha=-1.7499$ and $\beta=-2.7630 \cdot 10^{-2}$. The distribution of the residuals of the fitting in (\ref{eq:fit_log_ds_G}) is given in Table \ref{table:G_ds_measured}.  

\item Generate the values of $a_0$, $N$ and $A$ using the expressions given in Table \ref{table:fittings}.

\item Since $a_1$ appears to be uncorrelated with the remaining parameters of the model, it can be generated independently according to the distribution given in Table \ref{table:param}.
\end{enumerate}

Sections \ref{Statistical_Characterization} and \ref{relation_between_parameters} have revealed useful information about the probability distributions of $d_i$ and $g_i$ and their relation to other parameters of the \ac{MPM}. However, further research is required to generate  realistic sequences of these magnitudes. Our preliminary analysis suggests that the distribution of $d_i$ conditioned on $N$ can be modelled as a mixture of the Pearson and \ac{GEV} distributions. Nevertheless, a larger number of measured channels are required to obtain accurate estimates of the distribution parameters for high values of $N$. A mathematical fitting of the conditional \ac{PDF} of $|g_i|$ given $d_i$, illustrated in Fig. \ref{fig:3D_gi_vs_d_i_ds}, is also required. The generation of sequences of $d_i$ requires knowledge of  their joint distributions, which seems to be unrealistically complex. However, at least their covariance is required to obtain a linear model.  

\begin{figure}[h!]
	\centering
	\begin{subfloat}[]
		{\includegraphics[width=0.95\columnwidth]{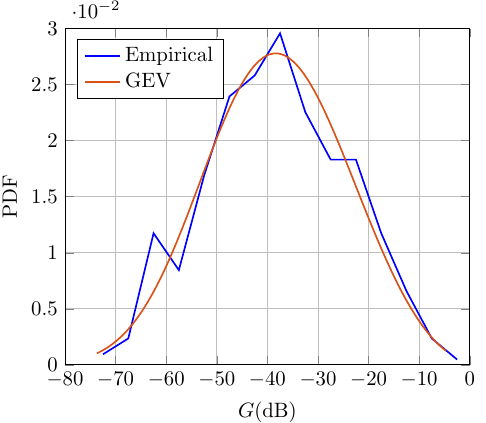}}
	\end{subfloat}
	\hfill
        \begin{subfloat}[]
		{\includegraphics[width=0.95\columnwidth]{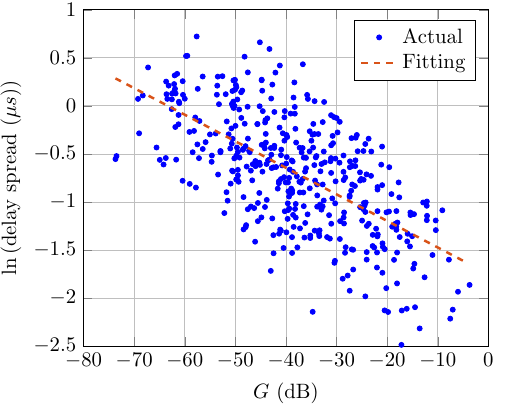}}
	\end{subfloat}
	\caption{{\color{reviewer1}(a) Empirical and fitted \acp{PDF} of the average channel gain; (b) scatter plot of the relation between the logarithm of the delay spread and the average channel gain.}}
	\label{fig:mesured_channels_charac}
\end{figure}

\begin{table}[hbt]
        \setlength{\tabcolsep}{3.2pt}
	\centering
	\caption{{\color{reviewer1}Fitting distributions of the average channel gain and of the residuals of the fitting in (\ref{eq:fit_log_ds_G})}}
{\color{reviewer1}
	\begin{tabular}{ccc}
		\toprule
		\begin{tabular}{c}\textbf{Channel}\\ \textbf{parameter}\end{tabular} & \textbf{Distribution} & \begin{tabular}{c}\textbf{Distribution}\\ \textbf{parameters}\end{tabular}\\
		\midrule
		\begin{tabular}{c}Average channel gain \\ $G$ (dB)\end{tabular} & GEV & 
		\begin{tabular}{c} 
			$k=-0.2984$\\ 
			$\sigma=13.9104$\\
			$\mu=-43.025$ 
		\end{tabular}\\
		\midrule
            \begin{tabular}{c}Residuals of the fitting in (\ref{eq:fit_log_ds_G})\end{tabular} & Normal & 
		\begin{tabular}{c} 
                $\mu=1.2854$\\
                $\sigma=0.3328$\\		
		\end{tabular}\\	            
		\bottomrule
	\end{tabular}
}
\label{table:G_ds_measured}
\end{table}
}

{\color{reviewer3}
\subsection{Relation between the \ac{MPM} parameters and the performance of indoor \ac{PLC} systems}
This section assesses the relation between the model parameters and the performance of a \ac{SISO} \ac{PLC} system like the one defined in the ITU-T Recommendation G.9960 \cite{G.9960}. To this end, a pulse-shaped \ac{OFDM} system with 4096 carriers in the frequency band up to $100$ MHz, a guard interval of $10.24$ $\mu$s and a rolloff interval of $5.12$ $\mu$s is employed. Constellations with up to $12$ bits/symbol, subject to an objective bit error rate of $10^{-6}$ are used. The \ac{PSD} defined in the ITU-T Recommendation G.9964 \cite{G.9964} is utilized, including the 20 subbands that are permanently excluded to avoid interfering existing wireless systems. As a result, the effective number of carriers in the band up to $f_{M-1}=80$ MHz is $3079$, which yields a maximum \ac{PHY} bit rate of $687$ Mbit/s. 

For the sake of simplicity, continuous \ac{OFDM} symbol transmission is assumed. A noise model with Gaussian distribution and \ac{PSD} given by $S_N(f)=a+bf^c$ (dBm/kHz), with $f$ in MHz, ${a= -1.7542 \cdot 10^3}$, ${b=1.6594 \cdot 10^3}$ and ${c=-4.1565 \cdot 10^{-3}}$ is used \cite{Cortes23}. 

Fig. \ref{fig:Bit_rate_vs_MPM} depicts the scatter plot of the \ac{PHY} bit rate versus three key parameters of the \ac{MPM}: the effective number of paths, $N$; the attenuation coefficient, $a_0$; and the normalization coefficient, $A$. As expected, performance decreases as $N$ and $a_0$ increases, in the former case because of the increased frequency selectivity of the channel and the latter because of the increased attenuation. As shown in Table \ref{table:bit_rate_fittings}, in both cases the relation can be fitted by means of a third-order polynomial function, although the cubic and quadratic terms have lower influence in the case of $a_0$, where the relation is almost linear. Regarding the relation to the normalization coefficient displayed in Fig. \ref{fig:Bit_rate_vs_MPM} (c), the performance increases with $A$. This is coherent with the fact that $A=1/\underset{i}{\max}(|g_i|)$, and the greater the absolute value of the path gains, the larger the frequency selectivity of the channel response.  

\begin{figure*}[ht]
	\centering
	\begin{subfloat}[]
		{\includegraphics[width=0.325\textwidth]{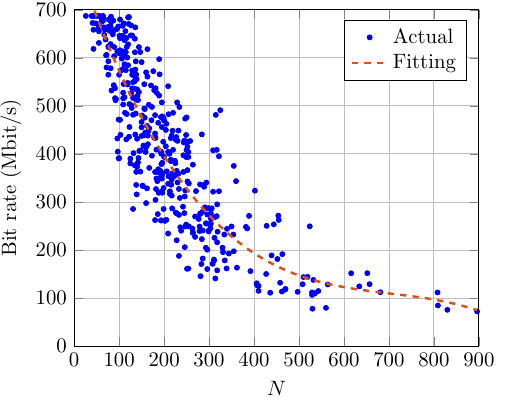}}
	\end{subfloat}
	\hfill
	\begin{subfloat}[]
		{\includegraphics[width=0.32\textwidth]{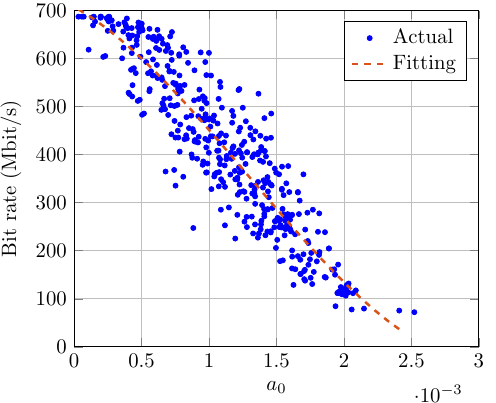}}
	\end{subfloat}
        \hfill
        \begin{subfloat}[]
		{\includegraphics[width=0.325\textwidth]{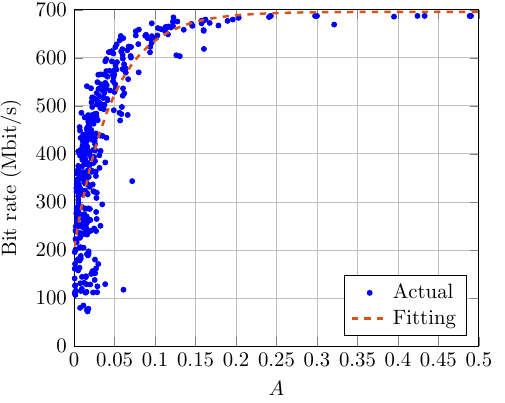}}
	\end{subfloat}
	\caption{{\color{reviewer3}Scatter plot and fitting curve of the estimated \ac{PHY} bit rate (Mbit/s) versus (a) the number of effective paths, $N$; (b) the attenuation coefficient $a_0$; (c) the normalization coefficient, $A$.}}
	\label{fig:Bit_rate_vs_MPM}
\end{figure*}

\newcolumntype{C}[1]{>{\centering\arraybackslash}m{#1}}
\begin{table*}[b]
	\centering
{\color{reviewer3}
	\caption{Parameters of the fitting expressions displayed in Fig. \ref{fig:Bit_rate_vs_MPM}.}
	\begin{tabular}{c C{2.5cm} C{2.5cm} C{2.5cm} C{2.5cm}}
		\toprule
		\textbf{Fitting expression} & $\bm{\alpha}$ & $\bm{\beta}$ & $\bm{\gamma}$ & $\bm{\delta}$\\
		\midrule
		$\widehat{R} \textrm{ (Mbit/s)}=\alpha + \beta \cdot N + \gamma \cdot N^2 + \delta \cdot N^3$ & $8.1687$ & $-2.7978$ & $3.8227\cdot 10^{-3}$ & $-1.8133\cdot 10^{-6}$\\
		\midrule
		$\widehat{R} \textrm{ (Mbit/s)}=\alpha + \beta \cdot a_0 + \gamma \cdot a_0^2 + \delta \cdot a_0^3$ & $7.0758 \cdot 10^{2}$ & $-1.5554\cdot 10^{5}$ & $-1.359\cdot 10^{8}$ & $3.5222\cdot 10^{10}$\\
		\midrule
		$\widehat{R} \textrm{ (Mbit/s)}=\alpha \cdot e^{\beta \cdot A}+\gamma$ & $-4.8969 \cdot 10^{-2}$ & $-21.0197$ & $6.9578 \cdot 10^{2}$ &-\\
        \bottomrule
	\end{tabular}
	\label{table:bit_rate_fittings}
}
\end{table*}

}
\section{Conclusion}
\label{sec:con}
 The \ac{MPM} is one of the most common top-down models adopted in the literature about \ac{PLC}. {\color{reviewer2}This paper explores the parameterization of this model, which in the \ac{PLC} context was firstly applied to low-voltage power distribution networks in the band up to 20 MHz}, for the indoor broadband scenario in the band up to 80 MHz. Since the number of parameters that result from the extension of the \ac{MPM} model to this context is extremely large, some of them have been imposed on the basis of common sense intuition, others, such as the attenuation parameters and the set of initial path lengths, have been chosen according to physical assumptions, and the remaining ones are analytically retrieved by means of a \ac{WLS} fitting. 
 
 Since the number of parameters that result after the \ac{WLS} fitting is still excessively large, only the dominant paths that actively contribute to synthesize the \ac{CFR} have been selected according to a proposed decimation procedure that yields a more compact model. The implemented selection method determines the number of significant paths subject to a constraint on the maximum value of the \ac{NRMSE}. 

 Once the \ac{MPM} parameters have been computed according to the proposed strategy, they are related to the characteristics of the actual \acp{CFR} like the frequency selectivity and the average channel gain. Suitable distributions that fit the empirical \ac{PDF} of each parameter are also given. The relation between the most relevant parameters of the model is explored. 
 
 {\color{reviewer1}Finally, the applicability of the provided results for the development of stochastic \ac{MPM} models for indoor broadband \ac{PLC} and the influence of the \ac{MPM} parameters on the performance of \ac{PLC} system are discussed.}

\bibliographystyle{IEEEtran}  
\bibliography{biblio}

\end{document}